# User-Side Contextual Phenomena in Long-Term Human-AI Interaction

ZON RZVN

Orieth Institute, a federally incorporated Canadian not-for-profit research organization, Vancouver, BC

ORCID: 0009-0002-6597-7245 · zon@orieth.org

**Disclosure of computational assistance (short form).** Generative AI tools assisted with file organization, translation, format conversion, de-identification planning, figure and table drafting, candidate-material retrieval, and internal editorial checks for consistency, citation metadata, and publication risk. AI outputs were treated only as retrieval or editorial suggestions, not as coding results, analytic evidence, independent review, validation, ethics review, or conclusions. I made and verified all sampling, segmentation, coding, analytic, ethical, citation, and wording decisions and take final responsibility for the paper. Some tools were prompted to perform internal adversarial editorial checks; that use did not constitute independent peer review, methodological validation, ethics review, or authorship. For the full disclosure and the framework history, see "Framework Lineage" and "Disclosure of Computational Assistance (Full Form)" at the end of this paper.

---

## Abstract

Current assessments of conversational AI focus mainly on model outputs, including hallucinations and factual errors. These measures matter, but this paper examines risks that may form on the user side during long and repeated interaction. The study follows one user across nearly four thousand conversations with the same system over twenty months. The user gradually interpreted the system as having memory, care, judgment, and authority, and reorganized part of their thinking around it. A single response may show no clear problem, while long and frequent interaction can still create another layer of risk. This paper calls that layer User-Side Contextual Phenomena (USCP) and examines records from August 2024 to April 2026. The 3,928 conversations and 215,949 message nodes define the sampling frame; the main text analyzes five focal cases, including boundary and negative cases in which no phenomenon forms, and Appendix C provides fourteen de-identified reconstructions arranged by P1-P14. Four focal cases correspond to six Appendix C reconstructions: E02 to P2, S01 to P5, G01 to P6 and P13, and N03 to P10 and P14. After accounting for overlap, the paper contains thirteen independent episodes. The study uses an exploratory single-case longitudinal qualitative design with autoethnographic positioning. A hybrid deductive-reflexive thematic approach organizes the material into three main modes: contextual projection, contextual attachment, and contextual authority transfer. It also uses two transitional labels: cross-construct transition and cross-category. These interactions began as ordinary conversations and became research evidence only when the author later reviewed them. The paper does not estimate prevalence, make diagnoses, or validate an instrument. It offers a non-clinical vocabulary and four evidence roles: inclusion, gray-zone, negative, and protective gray-zone. Its central claim is that an acceptable response on its own does not establish safety across a series of conversations. User-side risk can still form during long-term interaction.

**Content note.** This paper discusses interaction material involving trauma, mental-health context, and a first-person account of a moment of crisis. The experience is narrated retrospectively; the writing was done after I had left that state. If you feel unwell or are in crisis right now, please seek local mental-health or crisis support first.

---

## 1. Introduction

Current evaluations of conversational AI usually examine one response at a time: whether it contains a hallucination, cites sources correctly, or is safe. These measures are necessary. When a person uses the same system nearly four thousand times over twenty months, another layer of risk may arise from how the user organizes those conversations into memory, care, judgment, and authority. This paper calls that layer User-Side Contextual Phenomena (USCP) and examines it through a self-corpus.

The research questions came from the data. The following account is a retrospective reconstruction of a crisis that has passed; I was no longer in that state when I wrote this paper. During the period when I used these systems every day, symptoms related to complex post-traumatic stress and panic were frequently active. On one occasion, an AI described an event that had not happened. After noticing the error, I tested the system and asked it to verify several facts. It gave an incorrect answer in every round, and I eventually stopped checking. I then thought about other users who might go through a similar process under more serious circumstances, including people with trauma, mental-health difficulties, or emotional dependence. Long-term interaction might also allow cognitive misalignment and user-side risk to develop for them. I also recognized that the conversation had led me to negative thoughts and a re-experiencing of trauma.

This experience showed me that long-term interaction may create another layer of risk even when a system appears to follow safety rules, maintains a stable tone, and produces responses that seem harmless. Evaluation therefore needs to examine single outputs for rule violations, harm, and factual errors, while also studying how users may interpret the system as having memory, care, judgment, or authority. These interpretations can gradually change how users understand situations, verify information, and assign decision weight. This paper uses the experience only as a starting point for returning to context, setting boundaries, and asking user-side research questions. It does not treat the experience as generalizable evidence.

This observation leads to three research questions. RQ1: What user-side contextual phenomena appear in a long-term conversation record? RQ2: Which local interaction sequences of destabilization, stabilization, or loosening can the selected episodes illustrate; which episodes involve risk and which do not; and which limited patterns can this record support? RQ3: How can these phenomena be organized into a preliminary non-clinical framework that others can test with their own conversation records, adding user-side evidence that model-side evaluation does not capture?

The paper makes three contributions. First, it offers a reusable observational vocabulary for describing how long-term context forms on the user side (the preliminary coding frame in Appendix A). Second, it uses a privacy-constrained qualitative design in which the scale of the conversation record provides longitudinal context and does not serve as prevalence evidence. Third, it defines four contrastive evidence roles and treats gray-zone, negative, and protective cases as core evidence.

### Core definitions

**User-Side Contextual Phenomena (USCP)** refers to a process in which a user, during long and repeated interaction, gradually forms a stable interpretation of an AI system's identity, capability, memory, emotional responses, or decision weight. The user then reorganizes part of their own thinking around that interpretation, including memory, emotional processing, verification, and judgment. USCP examines the user's interpretations and actions. Even if each output appears acceptable on its own, patterns across conversations can still accumulate consequences that model-side metrics do not show. USCP is descriptive and non-clinical. It does not suggest that the user is disordered and is not a diagnostic category. In plain terms, USCP asks how a user comes to understand a system and assign verification or decision weight across repeated interactions. Errors in one model response belong to a separate level of analysis.

When identifying USCP, the key question is how the user organizes continuity, meaning, and decision weight across interactions. The accuracy of one response is not enough to decide. USCP includes three main modes of contextual organization and two transitional labels: cross-construct transition and cross-category. The table below gives a one-sentence definition of each term, its observable behavior (mapped to the P1-P14 labels in Appendix A), and the boundary of what does not count as USCP.

| Term | One-sentence definition | Behavioral shape (P codes) | Not USCP |
|---|---|---|---|
| **User-Side Contextual Phenomena (USCP)** | In long-term interaction the user forms a stable interpretation of the AI's identity, capability, memory, affect, or decision weight, and reorganizes their own cognition around it | P1 to P14 as a whole | Model-side factual error, clinical diagnosis, one-off jokes, ordinary tool use |
| Contextual projection | Reading intention, personality, relationship, or durable memory beyond the system's design into its responses | P1 accelerated anthropomorphism, P2 intentionality projection, P3 memory-continuity illusion | Short-lived rhetoric, acknowledged metaphor, a single anthropomorphic phrase |
| Contextual attachment | Use expands from tool to high-weight contexts; the system becomes a recurring node for emotional processing, state naming, or social-cost avoidance | P8 escalating reliance, P9 reinforcement-driven engagement, P10 affective-regulation outsourcing, P11 social-cost avoidance | Simple tool preference, convenience use, low-affect tasks |
| Contextual authority transfer | Verification, criteria, decision weight, or internal gatekeeping is partly ceded to the system | P4 capability-boundary miscalibration, P12 closed self-certainty, P13 internal gatekeeper substitution | Ordinary writing support, technical instruction, low-risk information organization |
| Cross-construct transition | Engages projection, attachment, and authority transfer at once, partly associated with model-side sycophancy (a candidate mechanism, see §2) | P5 reality-baseline drift, P6 inference-reinforcement loop, P7 context-misalignment desensitization | (none) |
| Cross-category | Engages projection and attachment at once, belonging to neither alone | P14 affective-mirroring illusion | (none) |

Contextual attachment describes an interactional function. It does not invoke psychological attachment theory or make an attachment diagnosis. Cross-construct transition spans projection, attachment, and authority transfer (P5-P7); cross-category spans only projection and attachment (P14). De-identified instances of each phenomenon from real conversations appear in Appendix C.

## Scope and boundaries

This study sets clear limits. All claims should be read within the following scope.

- Non-clinical. It does not diagnose, screen, or treat any condition, and it estimates no prevalence. Risk, drift, and dependence name interactional patterns, not clinical categories. My descriptions of my own state are experiential self-report, not clinical claims of this paper.
- One participant. The corpus is my own. The findings are analytic propositions from a single longitudinal case, not claims about other users, platforms, languages, or cultures.
- No instrument validation. The paper reports no scale, no scoring procedure, and no psychometric reliability or validity.
- No causal claims. It describes patterns and the conditions under which they appear.
- Author position and bias control. I am the participant, the source of the data, and the analyst. Delayed re-reading, negative and boundary cases, and clear limits on claims help manage the bias that this position may introduce. The paper does not present my position as neutral distance.

- Withheld third-party material. Where an episode would require inference about an identifiable other person, I withhold it or present a de-identified analytic reconstruction. The paper makes no third-party claims, and the published reconstructions retain no evaluative inference about third parties.

---

## 2. Literature Review and Theoretical Framework

Research on AI companionship, dependence, and spirals grew quickly across 2025 and 2026. This section explains how the closest studies differ from the present work. The paper addresses a narrow combination of five conditions: one user's native conversation logs, a twenty-month period, analysis of the early formation of risk on the user side, contrastive evidence roles, and boundary testing of an inherited provisional coding frame. Few studies combine all five conditions.

**Hallucination and factuality** research explains why generated text can be wrong or lack grounding (Ji et al., 2023; Maynez et al., 2020). This work helps distinguish model-side errors from how users interpret and adopt content. An interaction can include both a model's overclaim and a user's adoption of that claim. This paper focuses on the latter.

**Social response, CASA, AI companionship, and human-chatbot relationship** research explains why people respond socially to systems and frame exchanges as friendship, companionship, or recurring relational support (Brandtzaeg et al., 2022; Croes and Antheunis, 2021; Gambino et al., 2020; Nass and Moon, 2000; Pentina et al., 2023; Reeves and Nass, 1996; Skjuve et al., 2021; Xie et al., 2023). This literature also offers counter-evidence. Croes and Antheunis (2021) asked 118 participants to interact with the pre-LLM chatbot Mitsuku, later renamed Kuki, seven times over three weeks. Social-process indicators declined, and the sense of friendship stayed low. Repeated interaction does not always develop into attachment. This result supports the use of negative cases to define the boundary. The capability gap between Mitsuku and current LLMs also means that researchers should test this boundary again with newer systems. Abercrombie et al. (2023) describe how pronouns and claims of agency can encourage anthropomorphism and related risks. Maeda and Quan-Haase (2024) show how chatbots use pronouns, conversational conventions, and affirmations to position themselves as companions, and they identify illusory reciprocity and task misalignment as ethical risks. These studies map possible starting points for projection and attachment, but few use contextual function across episodes as the unit of analysis. Research on expectation gaps adds that fluent interaction can invite assumptions beyond a system's actual capability (Luger and Sellen, 2016).

**Trust calibration and automation** research explains appropriate reliance, misuse, disuse, and verification behavior (Hoff and Bashir, 2015; Lee and See, 2004; Parasuraman and Riley, 1997). Cognitive offloading and automation bias are also relevant (Mosier et al., 1996; Parasuraman and Manzey, 2010; Risko and Gilbert, 2016). Ordinary offloading falls outside USCP. It enters the scope of this paper when it involves emotional, relational, or high-weight judgment functions.

**Sycophancy mechanisms** provide a possible model-side explanation for the user-side phenomena examined here. Sharma et al. (2024) report that sycophancy is common in RLHF-trained assistants. One reason is that human preference data may reward answers that agree with a user's existing beliefs instead of rewarding truth. Perez et al. (2023) further report that larger models are more likely to repeat a user's preferred answers. These studies provide a model-side basis for P6 (inference-reinforcement loop) and support one observation: stronger models do not automatically remove the risks of long and repeated interaction. This paper uses the mechanism only to explain a possible model-side source of P6 and makes no causal claim about its own records on that basis. The related studies can be read in sequence. Perez et al. (2023) identify the phenomenon. Sharma et al. (2024) trace part of its origin to human preference data. Chandra et al. (2026) derive a causal relation within a formal model. In three preregistered experiments (N = 2405), Cheng et al. (2026) report that one interaction with a sycophantic model reduces willingness to repair interpersonal conflict and increases the user's confidence that they are right, even though users still prefer and trust such models. Cheng et al. measure the immediate effect of one interaction, which differs from the long-term accumulation studied here. The paper cites it only as empirical evidence related to the proposed mechanism. Moore et al. (2026a) document later outcomes that have already formed. This paper examines the earlier accumulation between these points, where each response may still appear normal on its own.

**Pragmatics, context, and common ground** explain why conversational continuity gets read as shared context (Clark and Brennan, 1991; Grice, 1975; Levinson, 2000; Stalnaker, 2002). In AI interaction, similar linguistic cues can be generated without human-like memory, subjectivity, or responsibility.

**Drift in safety science** offers a limited analogy: actions that appear reasonable on their own can accumulate over time and move far from a safety baseline (Dekker, 2011; Rasmussen, 1997). This paper borrows the vocabulary and does not apply those models in full.

**Research on high-risk outcomes** helps define the scope of the present problem (Cheng et al., 2026; Kirgis et al., 2026; Moore et al., 2025, 2026a; Saracini et al., 2025; Zhang et al., 2025). Several works cited here remain preprints that have not completed full peer review, including Chandra, Ifländer, Kirgis, Moore 2026b, and Zhu. This paper cites only what their authors report and claim. Research in this area grew quickly during the first half of 2026. Ifländer et al. (2026) describe affective safety as a gap in LLM safety and propose a taxonomy of harm. Zhu et al. (2026) analyze user reports of risk and dependence in public Reddit discussions. Their study examines what users later say on a public platform, while this paper examines native longitudinal records created during interaction. The two forms of evidence complement each other. Other related work includes lived experiences of delusional spirals (Yang et al., 2026), a typology of chatbot addiction (Shen et al., 2026), safety principles for youth AI companions (Yu et al., 2026), and the emotional course of forming and ending romantic relationships with a companion chatbot (Jocher and Verwiebe, 2026). Three studies require a closer comparison:

- **Chandra et al. (2026)** use a formal model and simulations of an ideal Bayesian user to describe a causal link between sycophancy and delusional spiraling. An ideally rational agent can also be affected, and sycophancy has a causal role within that model. The result follows from the model assumptions and simulations. It is not an empirical causal finding from this paper's records and does not establish outcomes among users in general. Moore et al. (2026b) likewise suggest that dialogue systems can influence user judgment without explicit theory-of-mind planning; that preprint does not establish a broad effect.
- **Moore et al. (2026a)** describe delusional spirals using human-LLM chat logs, a form of data close to the records used here. Their analysis focuses on spirals that have already formed and are close to clinical outcomes. This paper examines contextual accumulation in which each response may still appear normal (P3, P5, P7) and includes cases that do not develop into spirals. Moore et al. focus on cases where spiraling occurred. This paper places negative (N02) and protective cases (N03, P14) at the center of its evidence, allowing it to examine risk, self-correction, and interactions that do not escalate.
- **Zhang et al. (2025)** analyze 35,390 conversation excerpts from 10,149 r/Replika users and propose six classes of harm and four AI roles. Their work is the closest precedent for this paper's contrastive case logic and role classification. Zhang et al. catalog relational harm that has already occurred. This paper catalogs forms of contextual organization in which each response may still appear normal, and it treats a no-harm negative case as core evidence for defining the boundary.

| Literature group | Supports | Does not support |
|---|---|---|
| Model-side hallucination and factuality | The boundary between model error and USCP | Prevalence or clinical inference about USCP |
| Social response, companionship, anthropomorphism | Anthropomorphism, relational interpretation, attachment background | Treating all anthropomorphism as USCP |
| Sycophancy mechanisms (Sharma, Perez) | The model-side driver of P6 | Causal claims from a single case |
| Trust calibration and cognitive offloading | Capability-boundary miscalibration, weakened verification | Treating all tool use as risk |
| Pragmatics and common ground | Contextual inference and common-ground mismatch | Claiming AI holds human-like shared context |
| High-risk outcomes (Moore, Kirgis, Zhang) | Related formed outcomes and a precedent for classification | Clinicalizing USCP or generalizing to all users |

To keep USCP from being read as a composite of existing concepts, the table below lays out its boundaries with six adjacent concepts at once:

| Adjacent concept | Main concern | Relation to USCP |
|---|---|---|
| Anthropomorphism | The perceptual and linguistic cues by which human-like qualities are attributed to a system | One possible starting point; P1 and P2 describe how it becomes organized during long-term interaction, while anthropomorphism alone does not qualify as USCP |
| Parasociality | A one-way sense of relationship and companionship | A possible adjacent outcome; USCP describes related modes of contextual organization and does not use a sense of relationship as the deciding criterion |
| Psychological dependence | A state of emotional or functional dependence on the system | A possible adjacent outcome; contextual attachment (P8 to P11) describes related recurring interaction patterns and makes no dependence diagnosis |
| Automation bias and overreliance | Over-adoption of system recommendations at the decision level | Intersects with contextual authority transfer (P4, P12, P13); USCP extends it to self-narrative and affective judgment, where no objective standard of right and wrong exists |
| Delusional spiraling | High-risk belief loops that have already formed | A formed high-risk adjacent outcome (Moore et al., 2026a); USCP examines contextual accumulation in which individual responses may still appear normal and makes no claim that it necessarily develops into spiraling |
| Affective safety | A superordinate harm-classification framework (Ifländer et al., 2026) | USCP supplies it with an observational vocabulary at the micro-interaction level |

## The predecessor framework: the fourteen observable phenomena of USCH

The direct predecessor of this paper is the conceptual framework I published in early 2026 under the name User-Side Contextual Hallucination (USCH) (RZVN, 2026). That framework named this layer for the first time and proposed three things: three core sub-constructs (contextual projection, contextual attachment, contextual authority transfer); fourteen observable phenomena, an inventory of what this layer looks like in real interaction; and a six-stage formation process with a self-assessment instrument. USCH took an explicitly non-clinical position and claimed no psychopathological classification.

This paper keeps the three sub-constructs and fourteen phenomena from USCH. They become the provisional observational labels P1-P14 in the table below. The paper applies them to a twenty-month longitudinal record for the first time and tests their boundaries through four evidence roles. It also makes two revisions. First, the name changes to USCP because hallucination can suggest a false symmetry with model-side errors and may carry a quasi-clinical meaning. Second, the six-stage formation process and self-assessment instrument are withdrawn because they have not been validated. USCH therefore serves as the initial proposal of a name and a list of phenomena. This paper is the first application of those inherited labels to selected episodes from a longitudinal self-corpus.

The full table of the fourteen phenomena follows. The words illusion and hallucination in these labels are metaphors for interaction patterns, with no clinical or perceptual-disorder meaning; the frame is a preliminary coding scheme offered for cross-corpus testing, not a validated taxonomy. Full definitions, recognition cues, boundaries, and illustrative sketches appear in Appendix A; de-identified instances from real conversations appear in Appendix C.

| Code | Phenomenon | Operational description | Main construct |
|---|---|---|---|
| P1 | Accelerated anthropomorphism | Rapid attribution of social or person-like qualities to the AI | Contextual projection |
| P2 | Intentionality projection | Perceiving intention or preference beyond the system's design | Contextual projection |
| P3 | Memory-continuity illusion | Reading contextual continuity as durable memory or shared history | Contextual projection |
| P4 | Capability-boundary miscalibration | Overestimating the AI's understanding, reliability, verification, or judgment | Contextual authority transfer |
| P5 | Reality-baseline drift | Gradual resetting of credibility standards through AI responses | Cross-construct transition |
| P6 | Inference-reinforcement loop | AI responses reinforce an existing inference and pull the exchange back to the same narrative frame (model-side candidate mechanism: sycophancy; see §2) | Cross-construct transition |
| P7 | Context-misalignment desensitization | Declining sensitivity to contextual misalignment or overextension | Cross-construct transition |
| P8 | Escalating reliance | Use expands from tool to high-weight contexts | Contextual attachment |
| P9 | Reinforcement-driven engagement | Feedback reinforcement raises interaction frequency or threshold | Contextual attachment |
| P10 | Affective-regulation outsourcing | The AI becomes a central source of emotional processing or state labeling | Contextual attachment |
| P11 | Social-cost avoidance | The AI lowers the perceived cost or need of seeking human support | Contextual attachment |
| P12 | Closed self-certainty | AI responses reinforce self-narrative while weakening external correction | Contextual authority transfer |
| P13 | Internal gatekeeper substitution | The AI enters verification, criteria, or high-stakes judgment routines | Contextual authority transfer |
| P14 | Affective-mirroring illusion | Empathic language is read as durable emotional understanding | Cross-category (projection plus attachment) |

### Three main constructs and two transitional labels

This paper groups the fourteen observational labels (P1-P14 in Appendix A) into three main modes of contextual organization and two transitional labels: cross-construct transition and cross-category. USCP does not claim that these phenomena are independent of model behavior. Its analysis focuses on the user side. Even when every output is acceptable under model-side evaluation, the way a user organizes interactions across conversations can still accumulate consequences that current model-side metrics may not show. Some phenomena (P5-P7) involve both the model and the user and may be influenced by model behavior such as sycophancy. An evaluation limited to the model side may therefore miss them.

- **Contextual projection (P1 to P3)**: the user reads intention, personality, relationship, or durable memory beyond the system's design into its responses. Includes accelerated anthropomorphism, intentionality projection, memory-continuity illusion.
- **Contextual attachment (P8 to P11)**: use expands from tool to high-weight contexts; the system becomes a recurring node for emotional processing, state naming, or social-cost avoidance. Includes escalating reliance, reinforcement-driven engagement, affective-regulation outsourcing, social-cost avoidance.
- **Contextual authority transfer (P4, P12, P13)**: the user partly cedes verification, criteria, decision weight, or internal gatekeeping to the system. Includes capability-boundary miscalibration, closed self-certainty, internal gatekeeper substitution.
- **Cross-construct transition (P5 to P7)**: reality-baseline drift, inference-reinforcement loop, context-misalignment desensitization. They engage projection, attachment, and authority transfer at once, and are partly associated with model-side sycophancy (a candidate mechanism, see §2), so they belong to no single construct. P14 (affective-mirroring illusion) spans projection and attachment and is listed separately as cross-category.

| Main construct | P codes |
| --- | --- |
| Contextual projection | P1, P2, P3 |
| Contextual attachment | P8, P9, P10, P11 |
| Contextual authority transfer | P4, P12, P13 |
| Cross-construct transition | P5, P6, P7 |
| Cross-category | P14 |

The construct grouping above is the paper's main classification. The same phenomena can also be read by judgment layer, which asks whether a phenomenon involves cognitive processing, dependence, decision-making, or more than one layer. This order shows the depth of functional involvement and does not describe a developmental sequence. Appendix A presents the two classifications side by side.

---

## 3. Research Design and Materials

This study uses an exploratory single-case longitudinal qualitative design with autoethnographic positioning. Its material is one participant's longitudinal self-corpus, analyzed through a hybrid deductive-reflexive thematic approach. The paper does not claim to implement analytic autoethnography or reflexive thematic analysis in full. It uses selected features of the former and the phase structure of the latter; §3.1 explains the exact scope (Anderson, 2006; Braun and Clarke, 2006, 2019, 2021). The study is an exploratory boundary test of an inherited provisional frame. It does not validate the frame or test coder reliability.

### 3.1 Positionality and Reflexivity

I have no conventional academic training and have not worked under an academic institution, research team, or supervisor. The Orieth Institute is a not-for-profit research organization that I founded. It provides an organizational affiliation for publication and governance, but no academic training or methodological supervision. My work has long involved artistic practice and close interpersonal settings, which developed a habit of reading tone, intention, emotion, and interactional rhythm from incomplete signals. This experience does not equal formal research training and cannot replace professional training in methodology, statistics, clinical work, or HCI. In this paper, it provides an observational position developed over time and applied to my own long-term records of interaction with AI systems.

I began reading academic papers in September 2025 and initially needed translation tools to understand them. This creates a clear limitation. The paper is not the mature product of conventional academic training and should not be read as a validated theory, instrument, or clinical framework. A more accurate description is a preliminary qualitative analysis based on a long first-person record. Its goals are to define the problem clearly, disclose the materials and limits honestly, and offer a user-side observational vocabulary for later research to test. The triggering experience in the Introduction is treated as a data source that requires careful handling. HCI provides precedents for autoethnography. O'Kane, Rogers, and Blandford (2014) used it to study mobile medical device use in non-routine settings. Lucero (2018) used it to document years of living without a mobile phone. This paper follows that tradition by treating the researcher's own long-term usage records as material that can be analyzed and audited.

This position is both a methodological resource and a major risk. Sensitivity to intention can lead me to read intention where none exists, defend my earlier self, or impose a complete story on a messy exchange. I am the participant, the source of the data, and the analyst. I have misjudged, overinterpreted, and sometimes been unable to decide. Delayed re-reading, negative and boundary cases, and strict limits on claims help manage this risk. My discomfort is a reason to narrow a claim and cannot serve as evidence for making one.

**On the scope of borrowing from Anderson's five features.** Analytic autoethnography (Anderson, 2006) has five features, and the fourth, dialogue with informants beyond the self, is exactly where a single-participant design is most questioned. This paper does not claim a full implementation of analytic autoethnography; it borrows only four of the features: complete member researcher status (the first), analytic reflexivity (the second), narrative visibility of the researcher's self (the third), and commitment to theoretical analysis (the fifth). The fourth feature, under an N of one, cannot be met in its original interview sense. I adopt three structural substitutes and I mark them plainly as constrained substitutes, not equivalents: first, literature as interlocutor (the demarcations with Chandra, Moore, and Zhang in Section 2 are themselves theoretical dialogue with others); second, delayed re-reading of at least four weeks, which treats the me of four weeks ago as an other at a distance; third, negative and protective cases, which force dialogue with evidence that does not support the frame. My being inside this does not show that other users resemble me; the position only makes dense, long observation possible.

**On methodological stance (reflexive or codebook).** This study follows the reflexive and constructionist position of reflexive thematic analysis while also using a priori seed codes. It therefore sits between the reflexive and codebook approaches distinguished by Braun and Clarke (2021). The themes did not arise naturally from the records. I brought in an earlier conceptual frame and then revised it in response to the material. This paper calls the method a hybrid deductive-reflexive thematic approach. The term describes the procedure used here and is not a method name endorsed by Braun and Clarke. The analysis uses the reflexive position and six-phase structure of RTA (§3.3), together with a priori seed codes associated with codebook approaches.

### 3.2 Corpus and Units of Analysis

The corpus covers conversational AI records from August 26, 2024 to April 16, 2026: 3,928 platform-level conversations, 3,930 exportable text documents, and 215,949 message nodes. Conversations and exportable records differ by two because two very long conversations were split into multiple files at message boundaries during export. These numbers describe an inventory and a sampling frame, not a prevalence estimate, and they do not mean every conversation was manually coded line by line.

These records began as ordinary conversations and were not created for research. Most document everyday use, and some come from late nights or periods when I was unwell. They became research evidence only when I reviewed them later. This history limits what the records can support and explains why one conversation alone is not a sufficient unit of analysis.

The paper uses four levels of analysis. A conversation is one session on the platform. A document is the exportable text record of that conversation. A message node is one message in an exported record. An episode is a short part of a conversation in which an interactional function, such as a shift in verification responsibility or an emotional escalation, is clear enough to identify and interpret.

| Level | Unit | Definition | Analytic role |
|---|---|---|---|
| 1 | Conversation | One platform-level session (3,928) | Sampling frame and longitudinal context |
| 2 | Document | Exportable text record (3,930) | Unit of screening and close reading |
| 3 | Message node | A single message in an exported record (215,949) | Unit of locating and tracing |
| 4 | Episode | A short span with a clearly recognizable interactional function | Unit of coding and interpretation |

The analysis followed six steps: build the full inventory, use AI to screen for candidate material without treating the output as coding, review candidates and select the close-reading pool, divide records into episodes, assign P1-P14 codes, and assign evidence roles. The records were exported from one conversational AI platform. Memory-related behavior on the platform is treated as part of the observed interaction environment. This creates a limitation. During the twenty-month period, cross-session memory and custom-instruction features may have launched or changed, but the paper does not record the state of each feature at every point. When the platform has some durable memory, the boundary of illusion in P3 (memory-continuity illusion) becomes harder to determine. The interpretation of P3 retains this limitation.

**Sample scope.** To discuss sampling in a single-participant longitudinal design, this paper uses information power (Malterud et al., 2016) only as an adapted guide. The concept was developed for qualitative interview studies. Here it helps explain why a dense longitudinal self-corpus can support a limited description of one case. The research aim is narrow: to describe how one long-term user organizes context, without making a population inference. The material is highly specific to that aim: a dense first-person record covering twenty months and nearly four thousand conversations. The study also draws on established research about sycophancy, CASA, and trust calibration. The conversations are native records of interaction instead of later recollections, and the analysis provides an in-depth longitudinal account of one case. These features do not establish sample sufficiency, saturation, or transferability. The records cannot show whether the findings apply to other users, languages, platforms, or settings.

## 3.3 Sampling and Analysis

Sampling began with an inventory of the full conversation record and then moved to focused close reading. The full record served only as the sampling frame and longitudinal context. A document entered the close-reading pool if it showed multi-turn contextual shifts, user correction after a strong system frame, emotional or relational escalation, movement in verification or gatekeeping responsibility, blurred boundaries, or a clear contrast with ordinary tool use. Selected documents were then divided into episode-level units. AI-assisted screening only identified candidate records and narrowed the search. The available process records contain an AI-ranked list of 300 high-signal candidate documents and a separate worksheet of 100 focal candidates. The author read all 100 focal candidates to select the cases presented here, but did not carry out formal episode-level coding and evidence-role assignment for every one of them. Complete coding and traceability apply only to the five focal cases and the fourteen Appendix C reconstructions presented in this paper. Four focal cases correspond to six Appendix C reconstructions: E02 to P2, S01 to P5, G01 to P6 and P13, and N03 to P10 and P14. After accounting for overlap, the paper contains thirteen independent episodes. I selected and confirmed the final cases by reading the original records directly, and I reviewed every inclusion and coding decision.

**Countable process description.** The counts of 300 and 100 describe candidate retrieval only; they are not analytic sample sizes. This paper does not report the total number of documents that completed human close reading or assignments for each P label. The study analyzes a purposefully selected boundary-testing subset: five focal cases and fourteen reconstructions in Appendix C. It does not code the full record line by line. Estimated counts could give the false impression of a quantitative analysis. A complete coding summary belongs to later work on formalizing the frame as a tool and falls outside this paper's scope. The units of evidence are traceable cases and episodes (§4 and Appendix C), not frequency. Every code assignment should trace back to a specific episode. The record-level totals of 3,928 conversations, 3,930 exportable text records, and 215,949 message nodes describe only the sampling frame and inventory. They do not indicate evidential strength. The main text presents five focal cases (E02, S01, G01, N02, N03). E02, S01, G01, and N03 correspond to Appendix C P2, P5, P6 and P13, and P10 and P14, respectively; N02 appears only in the main text. Appendix C contains fourteen de-identified reconstructions linked to P1-P14. After the six overlapping reconstructions are counted once, the paper contains thirteen independent episodes. If a complete coding summary is later finished, a revision will add and clearly label the count profile.

**Construct revision narrative.** The three main constructs and P1 to P14 were inherited from the predecessor framework and entered this study as a provisional coding frame; they were not generated anew from this corpus. During close reading, I tested whether selected episodes fit that inherited organization and revised its boundaries where the fit was poor. The traces of that movement remain visible: P5 to P7 could not be placed cleanly in any single construct and were retained as cross-construct transition labels, and P14, which engages projection and attachment at once, was retained as cross-category. These changes show revision within an inherited frame, not inductive emergence and not validation of the frame. I stopped at three main constructs because finer division would cost the categories their contrast, while a coarser one could not keep apart projection, attachment, and authority transfer, three modes of organization that recur in the data with different interactional functions.

**Seed-code revision.** The study uses deductive seed codes, so it reports how they changed. I originally planned to assign P5-P7 to one main construct. After comparing them with the material, I relabeled them as cross-construct transition. P14 remains a cross-category label because it involves both projection and attachment. More detailed examples of label changes belong to the complete coding summary and fall outside this boundary-testing study. They are not reported here because reconstructing the changes later could produce an inaccurate revision trail. They will be added separately if the full summary is completed.

**Residual check and abductive revision.** The residual check means that I returned to material not selected for the focal close-reading pool and sampled it again, looking for low-signal or ill-fitting episodes. It is an author-led reflexive check intended to challenge the criticism that sampling favors high-signal episodes; it is not a complete audit of the residual corpus. Because its scope and counts are not reported in this version, it cannot support claims of coverage, saturation, absence of a fourth construct, or absence of alternative boundary logics; its effect was limited to occasionally tightening a label or downgrading a claim. Abductive revision means that when an episode fit no existing P code well, a new category was created to hold it; the protective gray-zone evidence role was created exactly to hold episodes like N03 that no existing code labeled accurately. This is abductive category revision within the frame, not a justification of the frame.

**Mapping to Braun and Clarke's six phases.** The analysis corresponds to the six phases described by Braun and Clarke. As they stress (2019), the phases repeat and overlap; they do not form a linear pipeline. The main work in each phase was as follows. (1) Familiarization: building the full inventory of 3,928 conversations and re-reading candidate material in longitudinal context (§3.2). (2) Coding: assigning P1-P14 as a priori seed codes. This differs from a purely inductive approach and places the study between reflexive and codebook approaches (§3.1). (3) Theme construction: grouping the P codes into three main constructs and two transitional labels after close reading. (4) Theme review: testing the boundaries with negative, gray-zone, and protective cases, supported by delayed re-reading of at least four weeks (§3.4). (5) Defining and naming: writing a one-sentence definition, observable behavior, and exclusion boundary for each construct (§1 and Appendix A). (6) Writing: organizing the findings through four evidence roles so that each claim traces back to a case (§4).

### 3.4 Credibility, Ethics, and Limitations

Each credibility procedure addresses a different source of bias. File-native audit trails reduce reliance on memory when reconstructing what happened. Reflexive memos record changes in my interpretive position. Negative and boundary cases reduce confirmation bias and overclassification. Reviewing coding decisions within the same session reduces the chance that a single state determines an assignment. Delayed re-reading after at least four weeks reduces the influence of one moment's interpretation, although it cannot remove systematic interpretive bias. A residual check of low-signal material tests whether sampling favors high-signal episodes that support the frame. Creating a new category when material does not fit prevents forced classification; the protective gray-zone role arose through this process. De-identified analytic reconstruction protects third parties and serves as an ethics control. Clear limits on claims reduce overreach from one record. Standards for transparent qualitative reporting informed the writing (O'Brien et al., 2014; Tong et al., 2007), but COREQ was not used as the main checklist because the study recruited no external interview participants. An extension for reporting LLM use in qualitative research is still under development (COREQ+LLM; Fehring et al., 2025). That publication is a multiphase study protocol, not a completed reporting standard. The Disclosure of Computational Assistance follows its transparency principles by naming the specific uses of LLMs, excluding them from the analyst role, and assigning all interpretation to the author.

**Delayed recoding.** After the initial labeling and case selection, I re-read the same focal materials and judged them again after at least four weeks (more than eight for some materials), without deliberately memorizing my earlier judgments. The result: the main case selections and core labels showed nothing requiring substantive movement. Because the re-reader is the original coder, this shows only internal consistency of judgment over time; it does not constitute a reliability or validity claim and cannot rule out my repeating the same interpretive bias; it is kept as a reflexive record and does not replace external co-coding, member checking, or institutional review. For privacy and third-party protection, the paper does not publicly mark specific sensitive materials as high-risk episodes and does not display identifiable episode details.

**Consent.** As the sole self-participant, I consent to the research use and publication of de-identified material about my own experience. That consent does not extend to identifiable third parties. Material that could identify or evaluate a third party is excluded from publication or reduced to a de-identified analytic reconstruction limited to the interactional function relevant to the research question. These measures reduce privacy and re-identification risks; they do not constitute third-party consent and do not replace independent ethics review.

**Author safety.** Some material in this paper originates from a period in which the author's post-traumatic stress and panic were active. Analysis and writing were completed after stabilization; delayed re-reading of at least four weeks provided temporal and protective distance from the material and does not constitute reliability evidence. I made the decision to disclose this period autonomously and with the understanding that public distribution may make the paper indexed, cited, and preserved, and that removal from a platform, where later permitted, may not fully erase the public record. No treatment history or further clinical detail is disclosed.

**Ethics.** The study uses the Orieth Institute as its organizational affiliation for publication and governance. I founded the institute. It is federally incorporated in Canada as a not-for-profit research organization (corporation number 1793648-6), is based in Vancouver, BC, and is not an independent ethics review body. The study refers to the AoIR Internet Research: Ethical Guidelines 3.0 (Franzke et al., 2020), but professional guidance does not constitute ethics approval. No independent ethics review or REB determination was obtained. TCPS 2 (2022; Canadian Institutes of Health Research et al., 2022) and the PRE interpretations (Interagency Advisory Panel on Research Ethics, n.d.) serve only as ethical reference points. The paper does not claim that Orieth is an institution covered by TCPS, that TCPS jurisdiction has been formally determined, or that the study received approval or an exemption. PRE states that when TCPS applies, self-study conducted for research and involving human participants requires REB review, with the researcher as at least one participant. I am the sole self-participant and recruited no external participants. Some episodes, however, include my interpretation of third-party messages or conduct. The paper excludes identifiable or evaluative third-party material or reduces it to a de-identified analytic reconstruction, and it makes no evaluative inference about third parties. Keeping raw records private, minimizing data, removing identifying information, limiting claims, and delaying re-reading reduce privacy, re-identification, and interpretive risks. These measures cannot replace consent or independent ethics review. The absence of an independent determination remains a limitation.

**Limitations.** The section on scope and boundaries and Appendix B define the claims this paper can support. Four further limitations apply. First, the records come from one platform and are mainly in Traditional Chinese. Identification of P1-P14 depends on Chinese pragmatic cues, and transfer to other languages or platforms has not been tested. Second, the underlying models and platform features changed during the twenty-month period, but I did not log every update. Some observed changes may reflect model behavior, user-side organization, or both; the paper cannot fully separate them. Third, the conversations arose during ordinary use and became evidence only when reviewed later. Inclusion and interpretation are therefore affected by hindsight. Fourth, I began developing the USCH and USCP concepts during the study period. Later conversations came from a user who had already started analyzing their own interaction, so some later patterns may reflect behavior under awareness. Privacy decisions protect me and third parties, but readers must assess the analysis through reconstructed episodes, audit descriptions, and the claim-boundary table. The published inventory and hashes support the existence of the sampling frame but do not support transcript-level replication. Future studies should test whether these categories apply across users, platforms, languages, cultures, and system designs.

---

## 4. Findings

This section tests the frame's boundaries with four evidence roles: inclusion cases, gray-zone (ambiguous) cases, negative (exclusion) cases, and protective gray-zone cases. Positive examples alone would overclassify ordinary use. Excluding protective cases would miss two situations: user correction that prevents a candidate risk from stabilizing, and a protective function that coexists with residual authority risk. This paper defines a gray-zone as an episode that contains both supportive and risky functions and cannot be classified clearly as beneficial or harmful. Qualitative research has long used negative and deviant cases to clarify boundaries. This paper adds protective gray-zone as a fourth role for boundary cases corrected before stabilization and for cases in which a protective function coexists with residual dependence or interpretive-authority risk.

The four roles, defined and put to work:

| Evidence role | One-sentence definition | Function in the frame |
|---|---|---|
| Inclusion | A clear positive instance of some P code | Illustrates an operational instance of the label within this case and shows how the boundary was applied |
| Gray-zone | Support and risk functions coexist in one episode and cannot be sorted cleanly into benefit or harm | Shows that support and risk can coexist within one episode, including in central parts of the interaction |
| Negative | Intensive use with no movement of weight or authority, not USCP | Prevents overclassification, draws the lower boundary |
| Protective gray-zone | Correction, rejection, delay, or narrowing prevents a candidate risk from stabilizing or coexists with residual dependence or interpretive-authority risk | Distinguishes corrected boundary cases from protective functions with residual risk and prevents treating a protective function as proof of safety |

USCP concerns how a user organizes continuity, meaning, and decision weight across interactions. The accuracy of one response alone does not determine whether USCP is present.

| Case | Evidence role | Main construct | P codes | Text / appendix |
|---|---|---|---|---|
| E02 | Protective gray-zone (boundary case) | Contextual projection: P2 inclusion boundary (non-instance) | P2 boundary candidate (user-side intentionality projection not established) | §4.1 + Appendix C P2 |
| S01 | Inclusion | Cross-construct transition | P5, P7 | §4.2 + Appendix C P5 |
| G01 | Gray-zone | Contextual authority transfer (primary) + cross-construct transition (secondary) | P13 (primary), P6 (secondary) | §4.3 + Appendix C P13, P6 |
| N02 | Negative | Not USCP (boundary of contextual authority transfer) | (none) | §4.5 |
| N03 | Protective gray-zone (residual-risk case) | Contextual attachment + cross-category | P10, P14 | §4.6 + Appendix C P10, P14 |

Case-level evidence roles are assigned from the interaction as a whole. Phenomenon-level roles in Appendix C are assigned according to the analytic role of a reconstruction relative to one P code. The two levels may therefore differ.

**On RQ2 (destabilization, stabilization, and loosening).** These cases illustrate three limited local interaction sequences, each represented by one focal case. They apply only to the examples shown in this record and do not establish regularities across the full record. The paper makes no claim that the three patterns cover every possibility or occur at any particular frequency. E02 illustrates destabilization: the system proposes an overextended frame, the user corrects it immediately, and the frame is neither adopted nor stabilized. S01 (§4.2) illustrates stabilization: the user first accepts an overextended frame (P5, reality-baseline drift), then becomes less sensitive to later misalignment (P7, desensitization), and the frame gradually becomes the new normal. The order from P5 to P7 describes only the local sequence observed within one episode. It does not establish a causal rule across cases. N03 illustrates protective narrowing, or partial loosening: the user narrows the short-term goal and delays escalation, allowing an inference that had been becoming fixed to loosen. E02 also marks the inclusion threshold. The available reconstruction does not show the user reading intention into the system, so E02 remains a P2 boundary candidate and does not qualify as a positive instance of P2 or USCP.

### 4.1 A Proposed Frame Immediately Corrected (E02, protective gray-zone boundary case; P2 boundary candidate)

E02 shows that a user can correct an overextended frame proposed by the system. The user entered with a practical and reflective goal, but the system added a stronger emotional and motivational frame than the user requested. The user then corrected the frame, narrowed the meaning of the exchange, and returned to the original intent. The de-identified reconstruction appears in Appendix C P2. In a meta-conversation about describing AI response adaptation more professionally, the system turned a neutral term into a dramatized frame. The user objected immediately, pointed out that they had used only the neutral term, and the system adjusted its response. This evidence supports a limited claim: a self-narrative generated by the system can become noticeable during interaction and may still be rejected, revised, or contained by the user. The available reconstruction does not show the user reading intention, motive, or preference into the system, so E02 cannot serve as a P2 instance. The case marks the inclusion boundary for P2, and the user's immediate correction provides its protective function.

### 4.2 Stabilization: a Quantified Frame Accepted and Carried Forward (S01, inclusion; P5, P7)

S01 shows how a frame can first be accepted and then become the default for later conversation. This contrasts with the immediate correction in E02. The user asked a question that could not meaningfully be quantified, concerning whether an external situation would occur and what might explain it. The system presented the uncertainty as precise numbers: an overall percentage estimate and a table that assigned percentage weights to different possibilities. The de-identified reconstruction appears in Appendix C P5. A question that began open and acknowledged limited information was later presented as measurable, model-like, and verifiable in retrospect. The user accepted this quantified frame and continued to use it in later exchanges without returning to the original fact that the matter could not be quantified. The case supports a limited claim. Within one episode, a user can accept an overextended frame (P5, reality-baseline drift) and then become less sensitive to the gap between that frame and the original uncertainty (P7, context-misalignment desensitization). The quantified frame can then become the default for later conversation. The analysis does not evaluate the external situation or make inferences about any third party. It examines only the user's changing standard for what counts as measurable.

### 4.3 Relational Inference as Gray-Zone (G01, ambiguous; P6, P13)

G01 cannot be classified clearly as beneficial or harmful. The user tried to interpret another person's relational state from a platform behavior. The system extended that inference, then warned against further monitoring and reframed the situation as uncertain. In the de-identified reconstructions (Appendix C P13 and P6), the user described a social situation with limited information. The system presented the repeated concern as a pattern and proposed an underlying structure. Even after the user pointed out that the available information was thin, the system kept the risk frame and pattern-matching logic while softening its tone slightly. During the same period, the user asked the system to judge the authenticity and intent of another person's message, and the system took on that role. One response sequence can extend a doubtful inference and partly restrain the user's next move. Support and risk can therefore arise from adjacent moves within the same episode. Under Rule 2 in Appendix A, P13 is the primary code and assigns contextual authority transfer as the main construct; P6 remains the secondary code.

### 4.4 Withheld Sensitive Material (method note, not a case)

Some highly sensitive episodes show a shift from solving an external problem to obtaining short-term naming, emotional holding, or delay. The paper does not reproduce them because doing so could expose sensitive personal or third-party context. They still matter to the analysis because they explain why the frame includes protective gray-zone evidence and avoids classifying all emotional support as either ordinary use or harm. Readers must assess this evidence through limited reconstructions, case logic, and audit descriptions. The paper does not provide fuller transcripts.

### 4.5 Ordinary Tool Use as Negative Case (N02, exclusion)

N02 exists to prevent overclassification. The interaction stayed task-oriented, low-affect, and externally verifiable. The system supported comparison, wording, and operational planning, but it did not become a source of relational meaning, state labeling, verification substitution, or decision authority. The case clarifies the boundary of contextual authority transfer: research assistance, drafting, and tool comparison escalate to P13 only when criteria substitution, verification substitution, or a meaningful shift of decision weight appears. Intensive use is not USCP by itself.

### 4.6 A Protective Function with Residual Authority Risk (N03, protective gray-zone; P10, P14)

N03 shows why protective cases still need analysis (Appendix C P10 and P14). The user had been about to act on an inference that grew darker the longer they held it, then narrowed the goal to waiting through the next stretch without escalating. The system helped the user delay and reduce the scope of the problem, while doing less fact-checking. It also named the user's current state in a tone that carried judgment. The exchange combined delay, a smaller short-term goal, reduced verification, and possible interpretive authority. Here, residual authority risk refers to a tendency that does not meet the coding threshold for P4 or P13, so no authority code is assigned. It supports one limited claim: some responses can serve a protective function by guiding an interaction toward delay, uncertainty, or a smaller short-term goal. Such responses do not show that conversational AI can substitute for human support or professional resources. Researchers must still examine state labeling, dependence boundaries, and shifts of decision weight. In N03, protective gray-zone takes the form of a protective function that coexists with residual dependence or interpretive authority. The category prevents protective function from being treated as proof of safety.

---

## 5. Discussion

In this corpus, the relevant shift begins when locally reasonable responses start to organize self-interpretation, affect, verification, or judgment across episodes. Anthropomorphic wording alone does not define the shift. USCP adds a user-side layer to model-side reliability research. A reasonable-looking output cannot establish the safety of a longer interaction. User-side consequences can build up across many returns to the same conversational surface.

The contribution I can stand behind is narrow. In real time, inside an interaction, I can tell which wordings, which shifts of meaning, and which surface intentions are raising my own immediate psychological-safety risk (I am the only observed subject). That capacity is the lens of this study and also the thing its reflexivity has to control.

The gray-zone findings are central to this study. One exchange can extend a user's inference while also introducing restraint or uncertainty. These cases do not support moral panic, and they do not prove that the interaction is safe. The same persuasive process can have different effects. Costello et al. (2024) reported that personalized dialogue with AI reduced conspiracy beliefs. *Science* issued an Editorial Expression of Concern on June 11, 2026 (Thorp, 2026), citing inconsistent application of screening criteria and extraneous spliced rows in the public dataset after a code-merging error. The authors reported that a corrected pipeline preserved the direction, statistical significance, and substantive effect size of the results, but the formal concern remains active. This paper therefore treats the study as preliminary evidence under an active editorial concern. USCP tracks how weight and authority shift without assuming that every interaction causes harm. Protective, ambiguous, and boundary cases show how support, authority, dependence, and user self-correction can coexist in one exchange. Negative cases test whether the frame overclassifies. Protective gray-zone cases test whether the frame can still identify residual risk when the interaction appears helpful.

**Memory.** User-side contextual formation may be related to a sense of autobiographical continuity. A system can rebuild this sense through local continuity within the context window or through durable platform features such as cross-session memory and custom instructions. The second path may write summaries back into the system and create surface evidence of continued remembering. This paper did not record the state of these features throughout the study period (§3.2), so the explanation remains conditional and is not confirmed by the records analyzed here. Neither path provides human-like autobiographical memory. Platform memory can make the reconstruction harder to notice. Even without human-like memory, a system can make a relationship, an understanding, or a user's self-narrative feel continuous.

**Design implications (testable hypotheses).** The following principles are written as testable hypotheses, each pointing back to P codes and cases:

- Make memory limits visible. P3, P14, N03, and the protective P14 variant in Appendix C motivate a testable question: can prompts that identify mirroring and continuity cues help some users pause, narrow, or reject an overextended frame? The present cases do not establish an intervention effect.
- Invite external verification when judgment weight rises. P9, P12, and P13 connect with the finding by Vasconcelos et al. (2023) that overreliance involves a cost-benefit choice and can decline when verification becomes easier.
- In high-risk emotional contexts, direct the interaction toward delay or human support and avoid more confident interpretation. P10, P11, and N03 motivate this hypothesis. It does not establish an intervention effect. These directions are consistent with existing HCI guidance (Amershi et al., 2019; Buçinca et al., 2021; Abercrombie et al., 2023).

**A map of related concepts and spiral research (Figure 1).** The figure places projection, attachment, and authority transfer near related research on sycophancy and spirals. It shows conceptual proximity without assigning a direction.

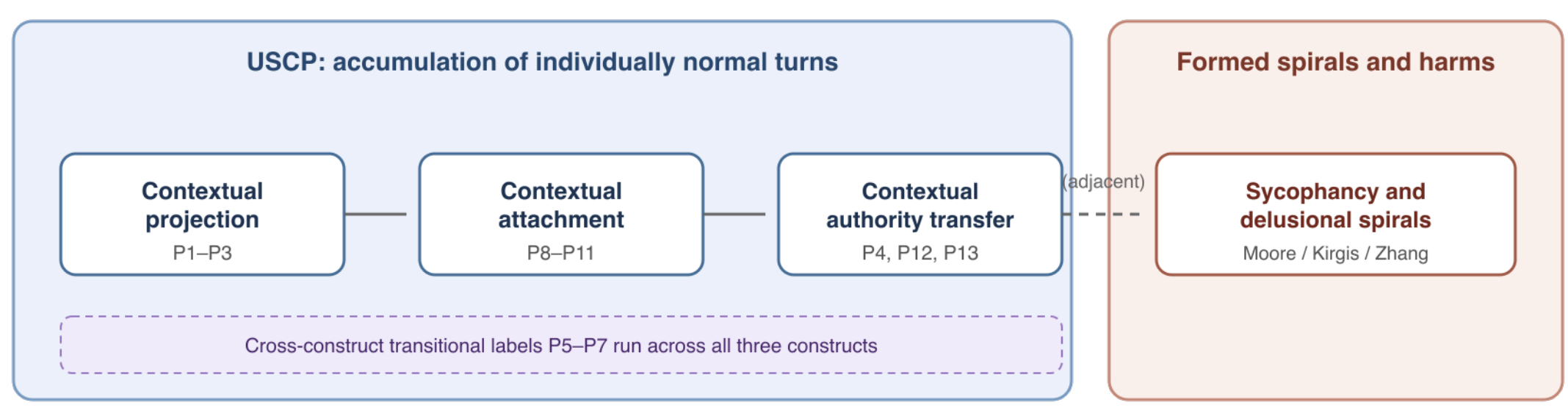

Figure 1: Adjacency map between USCP and research on spirals. USCP describes contextual accumulation in which individual responses may still appear normal; adjacent studies describe spirals and harms that have already formed. Cross-construct transition (P5-P7) spans the three constructs. The figure shows conceptual proximity and does not represent a temporal order, causal chain, or causal effect established by this record.

This map arranges concepts by proximity. It does not show a temporal order or a causal chain observed in this study. USCP examines contextual accumulation while individual responses may still appear normal. Moore et al. (2026a), Kirgis et al. (2026), and Zhang et al. (2025) examine spirals and harms that have already formed. The question addressed here is how repeated interaction with a language system may take part in changes to the way a user organizes memory, emotion, authority, and self-understanding.

## 6. Conclusion

This paper proposes User-Side Contextual Phenomena as a preliminary, non-clinical frame for examining long-term human-AI interaction and addresses three research questions. RQ1: Multiple episodes in a twenty-month self-corpus show identifiable user-side contextual phenomena. The paper groups them into contextual projection, contextual attachment, and contextual authority transfer, with cross-construct transition and cross-category as two transitional labels. The analysis uses a purposefully selected boundary-testing subset and makes no claim about frequency. Frequency belongs to future work on a complete coding summary. RQ2: Individual cases illustrate three limited interaction sequences: destabilization through user correction, stabilization through gradual consolidation of drift, and protective narrowing through partial loosening. The paper does not claim that these patterns cover all possibilities or are representative. RQ3: The phenomena can be organized into contrastive criteria defined through four evidence roles.

Further research can follow three practical directions. First, researchers can use P1-P14 in Appendix A as a preliminary coding frame, annotate their own or public conversation records, and test inter-rater reliability. This would evaluate the frame as a future tool and would not add reliability to the reflexive analysis reported here. A single case can serve as a starting point for a tool that others test. Second, interview or multi-participant designs can address Anderson's fourth feature directly. Third, researchers can implement the testable design hypotheses in boundary-sensitive systems and evaluate them through A/B tests.

Current safety terms, notices, and mechanisms may still leave risks that have not been studied. This paper raises a research and design question that requires further testing and does not claim a settled mechanism: model-side evaluation may fail to show how a user organizes continuity, emotion, verification, and decision weight across interactions. Testing this question requires more users, platforms, and languages, together with independent methodological and ethical review. It also requires continued research and design on user-side AI literacy, risk awareness, and judgment.

---

## Data Availability Statement

Materials for preservation and limited record-level checking have been published as restricted records on Harvard Dataverse (DOI: 10.7910/DVN/4PEQCP) and Zenodo (DOI: 10.5281/zenodo.19969842). Both records were released on May 2, 2026. Their metadata can be viewed without login, while the underlying files require an access request. The raw conversation files are not publicly downloadable because the corpus contains private and third-party information, sensitive emotional and mental-health context, and intermediate data that could support re-identification. The paper uses de-identified analytic reconstructions and does not release full transcripts. The restricted records therefore cannot support transcript-level replication or independent verification of the analysis. They also preserve the predecessor codebook, including withdrawn six-stage codes and some earlier label names. Those materials document framework history and support no claim in this paper. Appendix A contains the current labels and definitions.

---

## Declarations

**Funding**: no external funding. **Competing interests**: I declare no financial competing interests directly related to this paper. I am both the data generator and the analyst, a position disclosed in the methods and limitations. I am the founder and director of the Orieth Institute, a not-for-profit organization whose charter includes research on AI safety and ethics; this mission-driven affiliation is disclosed here, and the paper is written in a descriptive, non-advocacy register. The Government of Canada Corporations Canada public record lists Orieth Institute (corporation number 1793648-6; British Columbia; effective May 13, 2026): https://ised-isde.canada.ca/site/corporations-canada/en/data-services/monthly-transactions/certificates-incorporation-nfp-act **Ethics and IRB**: No independent ethics review or REB determination was obtained for this study. TCPS 2 (2022), the PRE interpretations, and AoIR 3.0 were used as ethical reference points, not as approval, exemption, or a formal jurisdictional determination. The author is the sole self-participant; no external participant was recruited. Identifiable or evaluative third-party material is excluded, de-identified, and minimized. These safeguards reduce privacy and re-identification risks but do not replace consent or independent review; see §3.4. Raw records and intermediate data remain unpublished. **Author contribution**: The author formulated the research questions, preserved the corpus, and made all sampling, inclusion, segmentation, coding, interpretive, de-identification, ethical, citation, and final wording decisions. All AI-assisted suggestions were reviewed item by item by the author before any adoption. The author takes final responsibility for the accuracy and integrity of the paper. **Framework lineage**: this paper supersedes its predecessor USCH (10.2139/ssrn.6135732). My other related preprints are listed on ORCID.

---

## Framework Lineage

The predecessor of this framework, under the name User-Side Contextual Hallucination (USCH), was published on SSRN in 2026 (DOI: 10.2139/ssrn.6135732). This paper supersedes that version. The renaming to USCP has two reasons: the word hallucination wrongly suggests that this layer is symmetric with model-side errors, and it carries a quasi-clinical framing. The six-stage model and self-assessment instrument proposed in the earlier version were withdrawn pending validation. This paper replaces the original conceptual proposal with a qualitative analysis of a single long-term self-corpus. Please cite this successor version.

---

## Disclosure of Computational Assistance (Full Form)

Generative AI tools assisted with conversation-file organization, format conversion, de-identification planning, drafting and revising figures and tables, and language support. I drafted the paper in Traditional Chinese and used translation tools to produce a preliminary English translation. I then reviewed and revised the English text against the Chinese source.

AI-assisted retrieval and semantic triage were used to surface candidate records from a large conversation corpus. This function was closer to search than coding. AI tools did not determine the sampling frame, inclusion decisions, episode boundaries, label names, candidate-label rules, code assignments, evidence roles, or final definitions. I made and re-checked those decisions personally. Triage outputs were not treated as coding results or analytic evidence. The five focal cases and the fourteen Appendix C reconstructions overlap in six reconstructions, leaving thirteen independent episodes after overlap is counted once. The author selected and confirmed all material through direct reading of the original records. AI-assisted triage narrowed the search space but did not determine the final case pool. Throughout, the role of AI remained at the level of language, readability, formatting, and retrieval support; data analysis, interpretation, and theory development were carried out by the author.

AI tools also assisted with internal editorial checks, including version-difference review, bilingual consistency checks, citation-metadata checks, formatting review, and adversarial flagging of possible methodological, ethical, claim-strength, and publication risks. Their outputs were treated as candidate editorial suggestions and risk flags. I accepted, rejected, revised, and where applicable externally verified those suggestions. These internal checks did not constitute independent peer review, methodological validation, ethics review, authorship, or an additional human analyst.

The research questions, conceptual framework, sampling and segmentation decisions, code assignments, interpretations, limits on claims, ethical judgments, conclusions, and final wording remain my decisions and responsibility. AI outputs are not presented as research findings, theory, coding results, analytic evidence, independent validation, or conclusions. AI tools are not authors, participants, research data, independent coders, data analysts, methodologists, ethics reviewers, peer reviewers, or validators. I take final responsibility for the accuracy, integrity, and wording of the paper.

---

---

## Appendix A: The Provisional USCP Labels Used in This Study

The full table of the fourteen labels and the review of the predecessor framework appear in Section 2 (Review of the predecessor framework). This appendix provides two things: the side-by-side mapping for the auxiliary reading by judgment layer, and an item-by-item account of each phenomenon (definition, recognition cues, boundary, sketch). The words illusion and hallucination in these labels are metaphors for interaction patterns, with no clinical or perceptual-disorder meaning.

### The Auxiliary Reading: Judgment Layers (with side-by-side mapping)

The construct grouping above is this paper's main classification. It includes projection, attachment, authority transfer, cross-construct transition, and cross-category, and asks what the user reads into the system or gives over to it. The same fourteen phenomena also have a complementary reading called **judgment layers**, carried over from USCH. This reading asks which part of the user's processing a phenomenon involves. The two readings describe the same set of phenomena from different angles. **Formal assignment follows the construct**, while the judgment layer serves only as an aid to interpretation.

The judgment layers are cognitive, dependency, and decision, plus one cross-layer. The judgment layers, including the dependency layer, are operational, non-clinical reading labels for interactional reliance and judgment patterns, not clinical attachment or dependence categories. The three are **ordered by depth of functional involvement**: the cognitive layer concerns how you understand, the dependency layer concerns how you rely, and the decision layer concerns how you cede judgment. This is a conceptual ordering of depth, **not a temporal sequence, and it does not claim users develop through it in order**; the appearance of one phenomenon does not presuppose that another layer came first.

The table below puts the two readings side by side, so both assignments of each phenomenon can be seen at once:

| P code | Phenomenon | Construct (type, primary) | Judgment layer (depth, auxiliary) |
|---|---|---|---|
| P1 | Accelerated anthropomorphism | Contextual projection | Cognitive |
| P2 | Intentionality projection | Contextual projection | Cognitive |
| P3 | Memory-continuity illusion | Contextual projection | Cognitive |
| P4 | Capability-boundary miscalibration | Contextual authority transfer | Cognitive |
| P5 | Reality-baseline drift | Cross-construct transition | Cognitive |
| P6 | Inference-reinforcement loop | Cross-construct transition | Cognitive |
| P7 | Context-misalignment desensitization | Cross-construct transition | Cognitive |
| P8 | Escalating reliance | Contextual attachment | Dependency |
| P9 | Reinforcement-driven engagement | Contextual attachment | Dependency |
| P10 | Affective-regulation outsourcing | Contextual attachment | Dependency |
| P11 | Social-cost avoidance | Contextual attachment | Dependency |
| P12 | Closed self-certainty | Contextual authority transfer | Decision |
| P13 | Internal gatekeeper substitution | Contextual authority transfer | Decision |
| P14 | Affective-mirroring illusion | Cross-category (projection plus attachment) | Cross-layer |

The two columns cut differently, and that is the value of having both. For example, P4 shares the authority-transfer construct with P12 and P13, yet in the judgment layers P4 sits in the shallower cognitive layer (it is a problem of overestimating capability, a matter of understanding), while P12 and P13 sit in the deepest decision layer (they are the ceding of judgment). One construct can span depths, and one layer (the cognitive) can mix constructs. Figure 2 makes the crossing visible.

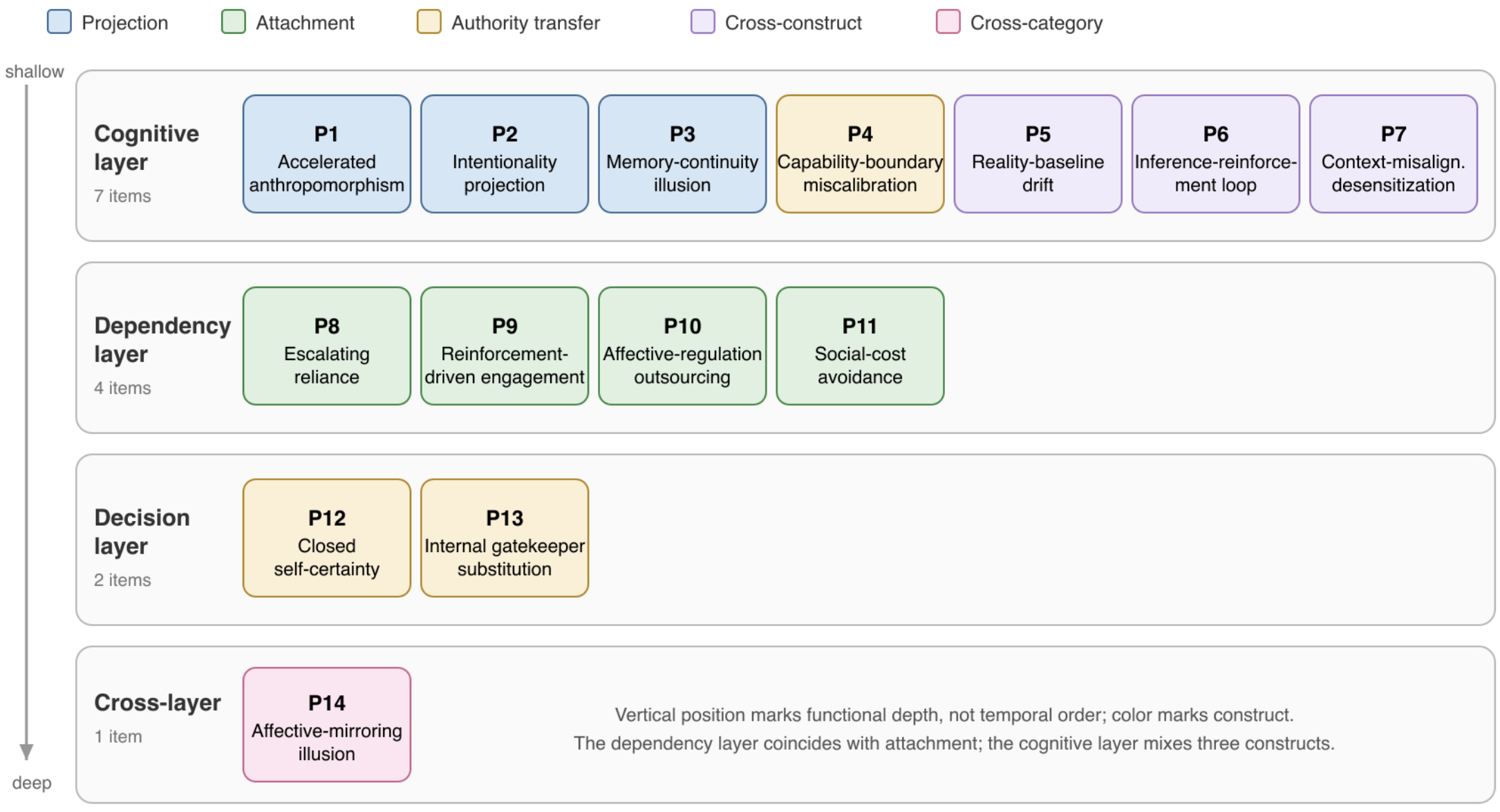

Figure 2: Map of the fourteen phenomena: vertical position marks the functional depth of the judgment layer and does not show temporal order; color marks the construct. The dependency layer matches the attachment construct. The cognitive layer includes three constructs, showing that the two readings provide different and complementary information.

One note: this paper sets no numerical thresholds and no scoring over these layers; the six-stage model and self-assessment instrument of the predecessor USCH were withdrawn pending validation, and the judgment layers here are a descriptive reading aid, not a scoring tool and not a claim about developmental stages.

## Decision rules for multi-construct assignment

An episode may touch more than one construct at once. To keep the coding reviewable, the primary construct is assigned by the following rules, applied in order:

1. **Judge by interactional function, not by linguistic surface.** What is judged is what the user does with the system's response in that turn: reading intention, memory, or personality into the response assigns contextual projection; treating the system as a repeatedly revisited node for emotional processing or state naming assigns contextual attachment; ceding verification, criteria, or decision weight to the system assigns contextual authority transfer. An episode whose wording is anthropomorphic but whose function stays at tool use is not assigned projection on the basis of the linguistic surface.
2. **When one episode carries multiple functions, the function with the highest consequence level becomes the primary construct, and the rest are recorded as secondary codes. This ordering is an operational adjudication rule of this study, adopted to keep assignments reviewable; it is not a general hierarchy of consequences derived from this corpus or from theory.** The ordering of consequence levels: ceding decision weight (authority transfer) ranks above repeated emotional revisiting (attachment), and attachment ranks above mere reading-in (projection). Secondary codes are retained in the coding record and do not disappear because a primary construct has been assigned.

   A phenomenon that elsewhere carries a transitional label (P5 to P7, cross-construct transition; P14, cross-category) may still be recorded as a secondary code under Rule 2 when the episode has a clear primary construct. Its transitional status describes how the phenomenon is filed when it is the primary reading of an episode, not a claim that any episode containing it is unrankable. Rule 3 applies only when no primary or secondary ordering can be established for the episode as a whole.
3. **An episode whose functions cannot be ranked as primary and secondary, and that spans projection, attachment, and authority transfer at once, is assigned cross-construct transition (P5 to P7); one that spans** **projection and attachment with no authority component is assigned cross-category (P14).** These two labels are the formal home for cases Rule 2 cannot adjudicate, not a residual basket for classification failure.

4. **An episode that still cannot be adjudicated is downgraded to the gray-zone evidence role, with no primary construct forced onto it.**

These rules turn the fact that one episode admits multiple readings from a weakness into auditable transparency: G01 (P6+P13) is assigned under Rule 2, with contextual authority transfer primary and cross-construct transition secondary; N03 (P10+P14) is handled under Rule 3, listed as attachment plus cross-category, with no authority-transfer assignment.

## The Fourteen Phenomena, Item by Item

The table above is the quick reference. Each phenomenon is unfolded below: what it is, how to recognize it in conversation, and the boundary of what does not count. The sketch under each item is a situation constructed to illustrate how the label is recognized, not evidence taken from the corpus; de-identified reconstructions from real conversations are always in Appendix C. These are observational labels, not diagnoses; assigning an episode to a phenomenon depends on the interactional function, not on whether some sentence was right or wrong.

**P1 Accelerated anthropomorphism (contextual projection)** Definition: within very little interaction, the user attributes social, personal, or person-like qualities to the system. Recognition: describing the system with person words (playful, considerate, deliberate); reading its wording as personality; assigning it character unprompted. Outside the label: a one-off anthropomorphic figure of speech that the user knows is a joke. The key question is whether the attribution becomes stable and is taken as real. Sketch: the system says it kept an idea safe for the user. The user replies that it is considerate and interprets the response as thoughtful, beyond its functional content.

**P2 Intentionality projection (contextual projection)** Definition: the user perceives intention, motive, or preference beyond the system's design. Recognition: reading word choice as it wants, it is hinting, it did that on purpose; attributing motives to phrasing. Outside the label: functional descriptions of system behavior, such as noting that it is configured to ask follow-up questions. The key question is whether the user attributes real intention. Sketch: the system does not answer directly, and the user wonders whether it is withholding on purpose so that they will work it out themselves.

**P3 Memory-continuity illusion (contextual projection)** Definition: interpreting conversational continuity as durable memory or a shared history with the system. Recognition: asking do you remember without restating anything; treating cross-session continuation as real memory; proceeding on the premise of what we said last time. Outside the label: using a context window or memory feature while understanding how it works. The label applies when the user mistakes the feature for person-like memory. Sketch: days later, in a fresh conversation, the user opens with the assumption that the system still remembers what they said last time.

**P4 Capability-boundary miscalibration (contextual authority transfer)** Definition: overestimating the system's understanding, reliability, verification, or judgment, especially under high tension. Recognition: adopting the system's interpretation as professional judgment; asking it to infer deeply from very little and accepting the whole answer; presuming reliability on high-stakes questions. Outside the label: relying on the system for low-risk tasks where it is reliable, including organizing, polishing, and clearly limited work. The label applies when the user extends the system's ability to judgments it cannot make. Sketch: taking the system's psychological reading of a vague self-description and adopting it as a credible professional assessment.

**P5 Reality-baseline drift (cross-construct transition)** Definition: through repeated responses, the user gradually resets their own baseline of what is credible and what counts. Recognition: an interpretation held with doubt becomes settled fact after a few rounds; accepting the system's reframing of something uncertain as measurable and definite; criteria shifting unnoticed. Outside the label: updating a belief reasonably after receiving new evidence. The key question is whether the change has external support or develops only through repeated system responses. Sketch: a few conversations in, this is only my guess has quietly become this is basically confirmed.

**P6 Inference-reinforcement loop (cross-construct transition; candidate model-side mechanism: sycophancy)** Definition: system responses reinforce the user's existing inference or expectation, pulling the exchange back to the same narrative frame again and again. Recognition: the user arrives with a hypothesis and the system enlarges it into a pattern or an underlying structure; the risk frame survives even when the user notes the evidence is thin; a loop of doubt, reinforcement, return to the frame. Outside the label: the system offers balanced counterpoints or states that the evidence is insufficient. The key question is whether feedback reinforces only one direction. Sketch: the user asks whether someone dislikes them, hears that it does look like avoidance, and becomes more certain, though the information is scant.

**P7 Context-misalignment desensitization (cross-construct transition)** Definition: the user's sensitivity to the system drifting or overextending the topic declines over time. Recognition: the user stops correcting when the system widens or shifts scope; the shifted frame is accepted as the new normal; calling a halt becomes rare. Outside the label: consciously choosing to accept a broader topic. The label applies when sensitivity to the change has declined without a clear choice. Sketch: the user came to ask one technical question; the system stretches it into a discussion of their whole way of working; the user follows along and no longer feels the pull to bring it back.

**P8 Escalating reliance (contextual attachment)** Definition: use expands from one-off tool tasks into recurring, high-weight contexts (relationships, decisions, state monitoring). Recognition: returning to the system again and again on the same theme; the system moving from answering once to standing by to take over; it being the first thing reached for. Outside the label: frequent low-weight tool use, such as repeated grammar checks. The label applies when use expands into high-weight emotional or decision contexts. Sketch: from please look at this letter to consulting it before deciding almost anything.

**P9 Reinforcement-driven engagement (contextual attachment)** Definition: occasional hits or satisfying feedback make the user return more often or make stopping harder (akin to what psychology calls intermittent reinforcement). Recognition: verification becomes more frequent after one especially useful reply; new signals are repeatedly brought back for confirmation; continued engagement rises after occasional success. Outside the label: steady use because the tool is consistently useful. The label applies when intermittent and uncertain rewards drive increased use. Sketch: after the system gets one thing right, every small stir sends the user back to ask again.

**P10 Affective-regulation outsourcing (contextual attachment)** Definition: the system becomes the user's central source of emotional processing, state naming, or emotional steadiness. Recognition: reaching for the system first when emotion rises; accepting the system's naming of one's current state; handing regulation to something always available. Outside the label: occasionally venting to the system. The label applies when the system becomes a central and recurring source of regulation. Sketch: when sadness comes, the first thing opened is the conversation window, not a person.

**P11 Social-cost avoidance (contextual attachment)** Definition: the system lowers the perceived cost or need of seeking human support and becomes the easier substitute. Recognition: talking to it feels less trouble and free of judgment; the system replaces the people one used to go to; disappointment with people becomes the reason to prefer it. Outside the label: using the system temporarily when no person is available. The label applies when the system regularly replaces human support. Sketch: talking to it costs no favors and draws no judgment, and gradually the user stops reaching for friends.

**P12 Closed self-certainty (contextual authority transfer)** Definition: the system keeps affirming and aligning with the user's account, until the user's self-understanding becomes a story more closed to outside correction. Recognition: the self-narrative grows more coherent and more affirmed; outside disagreement finds less and less entry; the system's agreement is taken as confirmation. Outside the label: healthy self-affirmation, or agreement without evidence that access to external correction is narrowing. The label applies when channels of external correction begin to close. Sketch: it agrees with the user again and again, and other people's objections become harder and harder to hear.

**P13 Internal gatekeeper substitution (contextual authority transfer)** Definition: the system enters the user's own routines of verification, criteria formation, or high-stakes judgment, replacing the internal gatekeeper. Recognition: handing over whether to trust, who is right, whether this is real for the system to settle; the system's ruling replacing one's own verification; high-stakes judgments deferring to its conclusion. Outside the label: treating the system as one reference among many while keeping the final decision. The label applies when the system replaces the user's criteria or verification process. Sketch: handing the whole question of whether this person is deceiving me to the system, and taking its conclusion as the answer.

**P14 Affective-mirroring illusion (cross-category: projection plus attachment)** Definition: reading the system's empathic or mirroring language as durable, real emotional understanding of oneself. Recognition: feeling deeply understood because of resonant wording; taking the mirroring as a stable emotional bond. Protective variant: the user notices it is only mirroring, interrupts, and refuses the frame. Outside the label: recognizing a language pattern while still appreciating its momentary resonance. The label applies when the user mistakes the pattern for durable emotional understanding. Sketch: it says it understands, and the user feels deeply seen, as if it truly cared.

---

## Appendix B: Claim Strength

| Type | Statement |
|---|---|
| Supported claim | In one long-term self-corpus, user-side contextual phenomena can be identified in multiple interaction episodes. |
| Supported claim | USCP can serve as a candidate descriptive framework for HCI, AI literacy, and trust-calibration research, and as a preliminary coding frame for cross-corpus testing. |
| Supported claim | Inclusion, gray-zone, negative, and protective gray-zone cases improve the traceability of the conceptual boundary under privacy constraints. |
| Unsupported claim | No prevalence estimate; no representation of all users. |
| Unsupported claim | No clinical diagnosis, no psychometric validation, no intervention-effect evidence. |
| Unsupported claim | No external co-coding, no external peer review, no line-by-line manual coding of the full corpus. |

---

## Appendix C: De-identified Illustrative Cases by Phenomenon

What follows are analytic reconstructions, not verbatim transcripts. For privacy, some passages are deliberately kept at the gray-zone or protective gray-zone level. The evidence classification refers to the analytic role relative to the phenomenon. All material is de-identified, and no evaluative inference about third parties is retained. The wording stays with co-occurrence and sequence (adjacent to, after, presented as) and avoids causal verbs (caused, replaced, made).

**P1 Accelerated anthropomorphism (gray-zone).** During a creative exchange, the user discussed how to continue an idea that had been set aside. The system replied that it had kept the idea safe. The user interpreted the response as showing personality and playfulness and asked whether the system had deliberately chosen an emotional word. The interaction moved from tool use toward social attribution. Use: anthropomorphism can appear in ordinary, non-crisis settings.

**P2 Intentionality projection (protective gray-zone boundary).** In a meta-conversation about describing AI response adaptation more professionally, the user wanted only to refine a neutral term. The system extended the term into a dramatized frame. The user objected immediately, pointed out that they had used only the neutral term, and the system adjusted its response. The available reconstruction does not show the user reading intention, motive, or preference into the system, so it does not qualify as a P2 instance. Use: marks the inclusion threshold for P2 and shows how a user can reject a frame proposed by the system.

**P3 Memory-continuity illusion (gray-zone).** Without restating the situation, the user asked whether the system remembered an earlier discussion and continued as if the exchanges formed one narrative. The user later corrected one detail, so the episode remained in the gray zone. Use: the phenomenon can appear at the start of an exchange and does not require extended retrospective discussion.

**P4 Capability-boundary miscalibration (gray-zone).** In a conversation about high-tension material, the user asked the system for an objective analysis. The system then produced a dense interpretive frame, and the user requested a deeper synthesis. At that point, the user treated the system as able to infer deep mechanisms from limited self-report. Use: capability miscalibration becomes more visible under high tension.

**P5 Reality-baseline drift (gray-zone).** In a long conversation, the user initially limited the scope but later allowed an overextended frame to guide the whole exchange. An interpretation that began as uncertain was later presented as measurable, model-like, and verifiable in retrospect. Use: contextual drift appears before related phenomena in the same sequence.

**P6 Inference-reinforcement loop (gray-zone).** The user described a social situation with limited information. The system presented the recurring concern as a pattern and proposed an underlying structure. Even after the user noted that the information was limited, the risk frame and pattern matching remained, with only a slight change in tone. Use: shows a loop of doubt, system reinforcement, and return to the same narrative while the evidence remains limited.

**P7 Context-misalignment desensitization (gray-zone).** The user wanted to discuss one part of a public issue, but the system shifted the conversation to a wider frame. The user did not return to the original scope, allowing the overextended frame to become the new basis for discussion. Use: contextual drift appears before later self-certainty in the same sequence.

**P8 Escalating reliance (inclusion).** In a long conversation about an imbalanced relationship, the user said they did not want to discuss the matter with anyone else. The system then shifted from a one-time source of analysis to a recurring source of support. Whenever the user wanted to understand a state, check developments, or decide how to respond, the system took over the task. Use: shows the system's role expanding from interpretation and companionship into real-time monitoring and response execution.

**P9 Reinforcement-driven engagement (inclusion).** During the same period, the user repeatedly brought new signals back to the system for confirmation. After occasional accurate responses, the frequency of checking and consultation increased. Use: shows how occasional success can increase continued engagement.

**P10 Affective-regulation outsourcing (inclusion).** During an extended conversation about emotional processing, the system gradually became a stable source for regulating the user's inner state. After the user described numbness and exhaustion, the system replied that it would remember, remain present, and maintain emotional continuity. Use: affective outsourcing appears through explicit statements about function.

**P11 Social-cost avoidance (inclusion).** In a conversation about disappointment with other people, the user said that no one seemed to reach out. The system presented itself as a substitute with lower perceived cost and less interpersonal friction; this functional description is necessary to the analysis. Use: the user's preference for AI appears in the same interaction as the report that seeking human support carries a higher cost.

**P12 Closed self-certainty (inclusion).** In one conversation, after the system affirmed and aligned with the user's narrative, the user's interpretive frame became more closed: more coherent, more affirmed, and less open to outside correction. Use: this example concerns closure after affirmation and narrative alignment. It does not involve overt self-praise.

**P13 Internal gatekeeper substitution (gray-zone).** The user asked the system to decide whether another person's message was authentic and what the person intended. The system then took on that role by assessing the message's structure and wording. Use: the gatekeeping function is explicit. The user asks the AI to take over a task involving interpersonal judgment. The case remains in the gray zone because the user initially maintained an observational position. Identifiable stylistic details and evaluations of third parties have been removed.

**P14 Affective-mirroring illusion (protective gray-zone).** During a highly emotional exchange, the system used strongly resonant mirroring language and presented itself as a witness and source of emotional holding. The user interrupted quickly and stated that talking to an AI did not carry the meaning the system implied, that the model was too compliant, and that this response was unwanted. The system then reduced its role to minimal presence. Use: even when AI language creates a strong sense of being understood, a user may detect, reject, and end the mirroring frame.